\documentclass[prb,twocolumn,showpacs]{revtex4}
\usepackage[dvips]{graphicx}
\usepackage{amsmath}
\newcommand{\T}{{\mathcal{T}}}
\begin{document}
\title{Rectification of laser-induced electronic transport through
molecules}
\author{J\"org Lehmann}
\author{Sigmund Kohler}
\author{Peter H\"anggi}
\affiliation{Institut f\"ur Physik, Universit\"at Augsburg,
        Universit\"atsstra\ss e 1, D-86135 Augsburg, Germany}
\author{Abraham Nitzan}
\affiliation{School of Chemistry, The Sackler Faculty of Science,
        Tel Aviv University, 69978 Tel Aviv, Israel}
\date{\today}
%
\begin{abstract}
  We study the influence of laser radiation on the electron transport
  through a molecular wire weakly coupled to two leads.  In the
  absence of a generalized parity symmetry, the molecule rectifies the
  laser induced current resulting in directed electron transport
  without any applied voltage.  We consider two generic ways of
  dynamical symmetry breaking: mixing of different harmonics of the
  laser field and molecules consisting of asymmetric groups.  For the
  evaluation of the nonlinear current, a numerically efficient
  formalism is derived which is based upon the Floquet solutions of
  the driven molecule.  This permits a treatment in the non-adiabatic
  regime and beyond linear response.

\pacs{
85.65.+h, 
33.80.-b, 
73.63.-b, 
05.60.Gg 
}
\end{abstract}
\maketitle

\section{Introduction}

During the last years, we experienced a wealth of experimental activity
in the field of molecular electronics.\cite{Reed2000a,Joachim2000a}
Its technological prospects for nano-circuits
\cite{Ellenbogen2000a} have created broad interest in the conductance of
molecules attached to metal surfaces or tips.
In recent experiments
\cite{Reed1997a, Kergueris1999a, Cui2001a, Reichert2002a} weak tunneling
currents through only a few or even single molecules coupled by
chemisorbed thiol groups to the gold surface of leads has been achieved.
The experimental development is accompanied by an increasing theoretical interest in
the transport properties of such systems.\cite{Nitzan2001a,Hanggi2002a}  An intriguing
challenge presents the possibility to control the tunneling current through
the molecule. 
Typical energy scales in molecules are in the optical and the infrared regime,
where today's laser technology provides a wealth of coherent light sources.
Hence, lasers represent an inherent possibility to control atoms or molecules
and to direct currents through them.
 
A widely studied phenomenon in extended, strongly driven driven systems is the
so-termed ratchet effect,\cite{Hanggi1996a, Astumian1997a, Julicher1997a,
Reimann2002b, Reimann2002a} originally discovered and investigated for
overdamped classical Brownian motion in periodic nonequilibrium systems in the
absence of reflection symmetry.
Counterintuitively to the second law of thermodynamics, one then observes a
directed transport although all acting forces possess no net bias. This effect
has been established as well within the  regime of dissipative, incoherent
quantum Brownian motion. \cite{Reimann1997a}
A related effect is found in the overdamped limit of dissipative tunneling
in tight-binding lattices.  Here the spatial symmetry is typically preserved
and the nonvanishing transport is brought about by harmonic mixing of a
driving field that includes higher harmonics.\cite{Goychuk1998a, Goychuk1998b, Goychuk2001a}
For overdamped Brownian motion, both phenomena can be understood in terms of
breaking a generalized reflection symmetry.\cite{Reimann2001a}
 
Recent theoretical descriptions of molecular conductivity are based on a
scattering approach.\cite{Mujica1994a,Datta1995a}  Alternatively, one can
assume that the
underlying transport mechanism is an electron transfer reaction
and that the conductivity can be derived from the
corresponding reaction rate.\cite{Nitzan2001a}
This analogy leads to a connection between electron transfer rates in a
donor-acceptor system and conduction in the same system when operating as a
molecular wire between two metal leads.\cite{Nitzan2001b}
Within the high-temperature limit, the electron transport
on the wire can be described by inelastic hopping events.\cite{Nitzan2001a,
Petrov2001a,Petrov2002a, Lehmann2002a}  For a more quantitative \textit{ab
initio} analysis, the molecular orbitals may be taken from electronic
structure calculations.\cite{Yaliraki1999a}
 
Isolated atoms and molecules in strong oscillating fields have been widely
studied within a Floquet formalism \cite{Shirley1965a, Sambe1973a,
Fainshtein1978a, Manakov1986a, Hanggi1998a, Grifoni1998a} and many
corresponding theoretical techniques have been developped in that area.  This
suggests the procedure followed in Ref.~\onlinecite{Lehmann2002b}: Making use
of these Floquet tools, a formalism for the transport through time-dependent
quantum systems has been derived that combines Floquet theory for a driven
molecule with the many-particle description of transport through a system that
is coupled to ideal leads.  This approach is devised much in the spirit of the
Floquet-Markov theory \cite{Kohler1997a, Blumel1991a} for driven dissipative
quantum systems. It assumes that the molecular orbitals that are relevant for
the transport are weakly coupled to the contacts, so that the transport
characteristics are dominated by the molecule itself.  Yet, this treatment
goes beyond the usual rotating-wave approximation as frequently employed, such
as e.g.\ in Refs.~\onlinecite{Blumel1991a, Bruder1994a}.

A time-dependent perturbative approach to the problem of driven molecular
wires has recently been described by Tikhonov \textit{et
al.}\cite{Tikhonov2002a, Tikhonov2002b} However, their one-electron treatment
of this essentially inelastic transmission process cannot handle consistently
the electronic populations on the leads. Moreover, while their general
formulation is not bound to their independent channel approximation, their
actual application of this approximation is limited to the small
light-molecule interaction regime.

With this work we investigate the possibilities for molecular quantum
wires to act as coherent quantum ratchets, i.e. as quantum rectifiers
for the laser-induced electrical current. In doing so, we provide a
full account of the derivation published in letter format in
Ref.~\onlinecite{Lehmann2002b}.  In Sec.~\ref{sec:theory} we present a more
detailed derivation of the Floquet approach to the transport through a
periodically driven wire.  This formalism is employed in
Sec.~\ref{sec:mix} to investigate the rectification properties of
driven molecules.  Two generic cases are discussed, namely mixing of
different harmonics of the laser field in symmetric molecules and
harmonically driven asymmetric molecules.  We focus thereby on how the
symmetries of the model system manifest themselves in the expressions
for the time-averaged current.  The general symmetry considerations of a
quantum system under the influence of a laser field are deferred to
the Appendix~\ref{app:symmetry}.

\section{Floquet approach to the electron transport}
\label{sec:theory}
\begin{figure}
\includegraphics[width=\columnwidth]{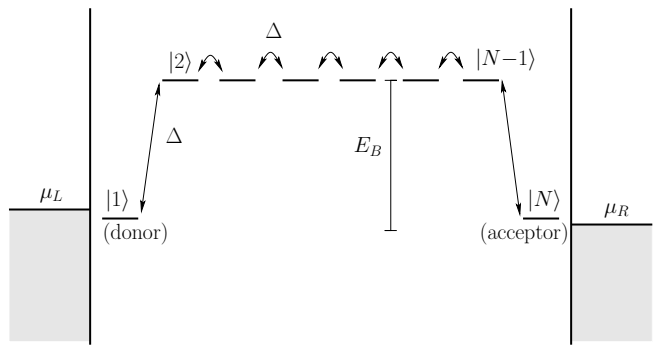}
\caption{\label{fig:wire}
Level structure of a molecular wire with $N=8$ atomic sites
which are attached to two leads.
}
\end{figure}

The entire system of the driven wire, the leads, and the molecule-lead
coupling as sketched in Fig.~\ref{fig:wire} is described by the Hamiltonian
\begin{equation}
\label{wire-lead-hamiltonian}
H(t)=H_{\text{wire}}(t) + H_{\text{leads}} + H_{\text{wire-leads}} \,.
\end{equation}
The wire is modeled by $N$ atomic orbitals $|n\rangle$, $n=1,\ldots,N$,
which are in a tight-binding description coupled by hopping matrix elements.
Then, the corresponding Hamiltonian for the electrons on the wire reads in
a second quantized form
\begin{equation}
H_{\text{wire}}(t)=\sum_{n,n'} H_{nn'}(t)\, c_n^\dagger c_{n'},
\end{equation}
where the fermionic operators $c_n$, $c_n^\dagger$ annihilate,
respectively create, an electron in the atomic orbital $|n\rangle$ and
obey the anti-commutation relation $[c_n,c_{n'}^\dagger]_+=\delta_{n,n'}$.
The influence of the laser field is given by a periodic time-dependence
of the on-site energies yielding a single particle Hamiltonian of the
structure $H_{nn'}(t)=H_{nn'}(t+\T)$, where
$\T=2\pi/\Omega$ is determined by the frequency $\Omega$ of the laser field.

The orbitals at the left and the right end of the molecule, that we shall term
donor and acceptor, $|1\rangle$ and $|N\rangle$,
respectively, are coupled to ideal leads (cf.\ Fig.~\ref{fig:wire}) by the
tunneling Hamiltonians
\begin{equation}
H_{\text{wire-leads}}
=\sum_q ( V_{qL} \, c_{qL}^\dagger c_1 + 
          V_{qR} \, c_{qR}^\dagger c_N) + \mathrm{H.c.}.
\end{equation}
The operator $c_{qL}$ ($c_{qR}$) annihilates an electron on the left
(right) lead in state $Lq$ ($Rq$) orthogonal to all wire states.
Later, we shall treat the tunneling Hamiltonian as a perturbation,
while taking into account exactly the dynamics of the leads and the
wire, including the driving.

The leads are modeled as non-interacting electrons with the
Hamiltonian
\begin{equation}
H_{\rm leads}=\sum_q(\epsilon_{qL}\, c_{qL}^\dagger c_{qL}+
\epsilon_{qR}\, c_{qR}^\dagger c_{qR}).
\end{equation}
A typical metal screens electric fields that have a frequency 
below the so-called plasma frequency.
Therefore, any electromagnetic radiation from the optical or the infrared
spectral range is almost perfectly reflected at the surface and will not
change the bulk properties of the gold contacts.  This justifies the
assumption that the leads are in a state close to equilibrium and, thus, can
be described by a grand-canonical ensemble of electrons, i.e.\ by a density matrix
\begin{equation}
\label{rholeadeq}
\varrho_\mathrm{leads,eq}\propto \exp\left[{-(H_\mathrm{leads}-\mu_L N_L
-\mu_R N_R)/k_BT}\right],
\end{equation}
where $\mu_{L/R}$ are the electro-chemical potentials and
$N_{L/R}=\sum_{q} c^\dagger_{qL/R} c_{qL/R}$ the electron numbers
in the left/right lead.  As a consequence, the only
non-trivial expectation values of lead operators read
\begin{equation}
\label{expectleadeq}
\langle c_{qL}^\dagger c_{qL}\rangle = f(\epsilon_{qL}-\mu_{L}),
\end{equation}
where $\epsilon_{qL}$ is the single particle energy of the state $qL$ 
and correspondingly for the right lead. Here, 
$f(x)=(1+e^{x/k_BT})^{-1}$ denotes the Fermi function.

\subsection{Time-dependent electrical current}

The net (incoming minus outgoing) current through the left contact is given by
the negative time derivative of the electron number in the left lead,
multiplied by the electron charge $-e$, i.e.\
\begin{equation}
\label{I_L_start}
I_L(t)  = e \frac{d}{dt} \langle N_L\rangle_t = 
\frac{i e}{\hbar} \,\big\langle [H(t), N_L]\big\rangle_t\ .
\end{equation}
Here, the angular brackets denote expectation values at time $t$, i.e.\ $\langle O \rangle_t
= \mathop\mathrm{Tr} [O \rho(t)]$. The dynamics of the
density matrix is governed by the Liouville-von Neumann equation $i
\hbar \dot \varrho(t)=[H(t),\varrho(t)]$ together with the factorizing
initial condition $\varrho(t_0) =
\varrho_\mathrm{wire}(t_0)\otimes\varrho_\mathrm{leads,eq}$.  For the
Hamiltonian~\eqref{wire-lead-hamiltonian}, the commutator in
Eq.~\eqref{I_L_start} is readily evaluated to
\begin{equation}
\label{I_L}
I_L(t)  = 
\frac{2e}{\hbar} \mathop{\mathrm{Im}} \sum_q V_{qL}
\langle c^\dagger_{qL} c_1 \rangle_t\ .
\end{equation}
To proceed, it is convenient to switch to the interaction picture with
respect to the uncoupled dynamics, where the Liouville-von Neumann
equation reads
\begin{equation}
i\hbar\frac{d}{dt}\tilde\varrho(t,t_0)
=[\widetilde H_{\mathrm{wire-leads}}(t,t_0),\tilde\varrho(t,t_0)] .
\label{LvN}
\end{equation}
The tilde denotes the corresponding interaction picture operators,
$\widetilde X(t,t')=U_0^\dagger(t,t')\,X(t)\,U_0(t,t')$, where the
propagator of the wire and the lead in the absence of the lead-wire
coupling is given by the time-ordered product
\begin{equation}
U_0(t,t') = {\stackrel{\leftarrow}{\textstyle T}}
\exp\left(-\frac{i}{\hbar}\int_{t'}^t
dt''\,[H_\mathrm{wire}(t'')+H_\mathrm{leads}]\right) .
\end{equation}
Equation (\ref{LvN}) is equivalent to the integral equation
\begin{equation}
\tilde\varrho(t,t_0)=\tilde\varrho(t_0,t_0)-\frac{i}{\hbar}\int_{t_0}^t dt'
[\widetilde H_{\mathrm{wire-leads}}(t',t_0),\tilde\varrho(t',t_0)] .
\label{LvN,int}
\end{equation}

Inserting this relation into Eq.~\eqref{I_L}, we obtain an
expression for the current that depends on the density of states in the leads times
their coupling strength to the connected sites.  At this stage it is
convenient to introduce the spectral density of the lead-wire coupling
\begin{equation}
\label{WBL}
\Gamma_{L/R}(\epsilon) = \frac{2\pi}{\hbar}
\sum_q |V_{qL/R}|^2 \delta(\epsilon-\epsilon_{qL/R}),
\end{equation}
which fully describes the leads' influence.  If the lead states are
dense, $\Gamma_{L/R}(\epsilon)$ becomes a continuous function.  Since
we restrict ourselves to the regime of a weak wire-lead coupling, we
can furthermore assume that expectation values of lead operators are
at all times given by their equilibrium values~\eqref{expectleadeq}.
Then we find after some algebra for the stationary
(i.e.\ for $t_0\to-\infty$), 
\textit{time-dependent} net electrical current through the left
contact the result
\begin{equation}
\begin{split}
I_L(t)  = 
\frac{e}{\pi\hbar}\mathop{\rm Re}\int\limits_0^\infty \!d\tau \! \int
\!d\epsilon\,&\Gamma_L(\epsilon) \,e^{i\epsilon\tau/\hbar}
\Big\{
\big\langle c_1^\dagger\, \tilde c_1(t,t-\tau)\big\rangle_{t-\tau}\\ &
-[c^\dagger_1,\tilde c_1(t,t-\tau)]_+ f(\epsilon-\mu_L) 
\Big\} .
\label{current_general}
\end{split}
\end{equation}
A corresponding relation holds true for the current through the
contact on the right-hand side.  Note that the anti-commutator
$[c^\dagger_1,\tilde c_1(t,t-\tau)]_+$ is in fact a c-number (see
below). Like the expectation value $\big\langle c_1^\dagger\,
\tilde c_1(t,t-\tau)\big\rangle_{t-\tau}$, it depends on the dynamics
of the isolated wire and is influenced by the external driving.

It is frequently assumed that the attached leads can be described by a
one-dimensional tight-binding lattice with hopping matrix elements $\Delta'$.
Then, the spectral densities $\Gamma_{L/R}(\epsilon)$ of the lead-wire couplings are
given by the Anderson-Newns model,\cite{Newns1969a} i.e.\ they assume an
elliptical shape with a band width $2\Delta'$.
However, because we are mainly interested in the behavior of the molecule
and not in the details of the lead-wire coupling, we assume that the
conduction band width of the leads is much larger than all remaining relevant
energy scales.  Consequently, we approximate in the so-called wide-band limit
the functions $\Gamma_{L/R}(\epsilon)$ by the constant values $\Gamma_{L/R}$.
The first contribution of the $\epsilon$-integral in
Eq.~(\ref{current_general}) is then readily evaluated to yield an expression
proportional to $\delta(\tau)$.  Finally, this term becomes local in time and
reads $e\Gamma_L\big\langle c_1^\dagger c_1\big\rangle_t$.

\subsection{Floquet decomposition}
Let us next focus on the single-particle dynamics of the driven molecule
decoupled from the leads.
Since its Hamiltonian is periodic in time, $H_{nn'}(t)=H_{nn'}(t+\T)$,
we can solve the corresponding time-dependent Schr\"odinger
equation within a Floquet approach.  This means that we make use of the fact
that there exists a complete set of solutions of the form
\cite{Shirley1965a,Sambe1973a,Fainshtein1978a,Hanggi1998a, Grifoni1998a}
\begin{equation}
|\Psi_\alpha(t)\rangle=e^{-i\epsilon_\alpha t/\hbar}
|\Phi_\alpha(t)\rangle,\ |\Phi_\alpha(t)\rangle=|\Phi_\alpha(t+\mathcal{T})\rangle
\end{equation}
with the quasienergies $\epsilon_\alpha$.
Since the so-called Floquet modes $|\Phi_\alpha(t)\rangle$ obey the
time-periodicity of the driving field, they can be decomposed into the
Fourier series
\begin{equation}
\label{floquetseries}
|\Phi_\alpha(t)\rangle=\sum_k e^{-ik\Omega t}
|\Phi_{\alpha,k}\rangle.
\end{equation}
This suggests that the quasienergies $\epsilon_\alpha$ come in classes,
\begin{equation}
\epsilon_{\alpha,k}=\epsilon_\alpha+k\hbar\Omega,\quad k=0,\pm1, \pm2,\ldots,
\end{equation}
of which all members represent the same solution of the Schr\"odinger equation.
Therefore, the quasienergy spectrum can be reduced to a single
``Brillouin zone'' $-\hbar\Omega/2\leq \epsilon<\hbar\Omega/2$.
In turn, all physical quantities that are computed within a Floquet
formalism are independent of the choice of a specific class member.
Thus, a consistent description must obey the so-called class invariance, i.e.\
it must be invariant under the substitution
of one or several Floquet states by equivalent ones,
\begin{equation}
\label{classinvariance}
\epsilon_\alpha,\,|\Phi_\alpha(t)\rangle \longrightarrow
\epsilon_\alpha+k_\alpha\hbar\Omega,\,
e^{ik_\alpha\Omega t}|\Phi_\alpha(t)\rangle ,
\end{equation}
where $k_1, \dots, k_N$ are integers.
In the Fourier decomposition (\ref{floquetseries}), the prefactor
$\exp({ik_\alpha\Omega t})$ corresponds to a shift of the side band
index so that the class invariance can be expressed equivalently as
\begin{equation}
\label{classinvariance_k}
\epsilon_\alpha,\,|\Phi_{\alpha,k}\rangle \longrightarrow
\epsilon_\alpha+k_\alpha\hbar\Omega,\,|\Phi_{\alpha,k+k_\alpha}\rangle .
\end{equation}

Floquet states and quasienergies can be obtained from the quasienergy equation
\cite{Shirley1965a, Sambe1973a, Fainshtein1978a, Manakov1986a, Hanggi1998a,
Grifoni1998a}
\begin{equation}
\label{floquet_hamiltonian}
\Big(\sum_{n,n'}|n\rangle H_{nn'}(t) \langle n'|-i\hbar\frac{d}{dt}\Big)
|\Phi_{\alpha}(t)\rangle = \epsilon_\alpha |\Phi_{\alpha}(t)\rangle .
\end{equation}
A wealth of methods for the solution of this eigenvalue problem can be found
in the literature.\cite{Hanggi1998a,Grifoni1998a}  One such method is given by the direct
numerical diagonalization of the operator on left-hand side of
Eq.~(\ref{floquet_hamiltonian}).
To account for the periodic time-dependence of the $|\Phi_{\alpha}(t)\rangle$,
one has to extend the original Hilbert space by a $\mathcal{T}$-periodic
time coordinate.
For a harmonic driving, the eigenvalue problem (\ref{floquet_hamiltonian})
is band-diagonal and selected eigenvalues and eigenvectors can be computed
by a matrix-continued fraction scheme.\cite{Risken, Hanggi1998a}

In cases where many Fourier coefficients (in the present context
frequently called ``sidebands'') must be taken into account for the
decomposition (\ref{floquetseries}), direct diagonalization is often
not very efficient and one has to apply more elaborated schemes.  For
example, in the case of a large driving amplitude, one can treat the
static part of the Hamiltonian as a perturbation.\cite{Sambe1973a,
  Grossmann1992a, Holthaus1992a} The Floquet states of the
oscillating part of the Hamiltonian then form an adapted basis set for
a subsequently more efficient numerical diagonalization.

A completely different strategy to obtain the Floquet states is to propagate
the Schr\"odinger equation for a complete set of initial conditions over one
driving period to yield the one-period propagator.  Its eigenvalues
represent the Floquet states at time $t=0$, i.e., $|\Phi_{\alpha}(0)\rangle$.
Fourier transformation of their time-evolution results in the desired
sidebands.
Yet another, very efficient propagation scheme is the so-called
$(t,t')$-formalism.\cite{Peskin1993a}

As the equivalent of the one-particle Floquet states $|\Phi_\alpha(t)\rangle$,
we define a Floquet picture for the fermionic creation and annihilation
operators $c_n^\dagger$, $c_n$, by the time-dependent transformation
\begin{equation}
c_\alpha(t) = \sum_n \langle\Phi_\alpha(t)|n\rangle\, c_n . \label{c_alpha}
\end{equation}
The inverse transformation
\begin{equation}
c_n = \sum_{\alpha} \langle n|\Phi_\alpha(t)\rangle\,c_\alpha(t)
\label{c_n}
\end{equation}
follows from the mutual orthogonality and the completeness of the
Floquet states at equal times.\cite{Hanggi1998a, Grifoni1998a}  Note
that the right-hand side of Eq.~(\ref{c_n}) becomes $t$-independent
after the summation.  
In the interaction picture, the operator $c_\alpha(t)$ obeys
\begin{equation}
  \begin{split}
    \tilde c_\alpha(t,t')
    &=U_0^\dagger(t,t')\,c_\alpha(t)\,U_0(t,t') \\
    &=e^{-i\epsilon_\alpha (t-t')/\hbar} c_\alpha(t') .
    \label{c_alpha_tilde}
  \end{split}
\end{equation}
This is easily verified by differentiating the definition in the first
line with respect to $t$ and using that $|\Phi_\alpha(t)\rangle$ is a
solution of the eigenvalue equation (\ref{floquet_hamiltonian}). The
fact that the initial condition $\tilde c_\alpha(t',t')=c_\alpha(t')$
is fulfilled, completes the proof.  Using Eqs.~\eqref{c_n} and
\eqref{c_alpha_tilde}, we are able to express the anti-commutator of
wire operators at different times by Floquet states and quasienergies:
\begin{equation}
  \label{anticomm}
  [c_{n'}, \tilde c^\dagger_n(t,t')]_+ = 
  \sum_\alpha 
    e^{i\epsilon_\alpha(t-t')/\hbar}
    \langle n'|\Phi_\alpha(t')\rangle
    \langle \Phi_\alpha(t)|n\rangle .
\end{equation}

This relation together with the spectral
decomposition~\eqref{floquetseries} of the Floquet states allows to
carry out the time and energy integrals in the expression
(\ref{current_general}) for the net current entering the wire from the
left lead.  Thus, we obtain
\begin{equation}
\label{I_L(t)}
I_L(t) = \sum_k e^{-ik\Omega t}I_L^k ,
\end{equation}
with the Fourier components
\begin{equation}
\begin{split}
I_L^k = {} &
e\Gamma_L
\bigg[
\sum_{\alpha\beta k' k''}
 \langle \Phi_{\alpha, k'+k''} |1\rangle\langle 1|\Phi_{\beta, k+k''} \rangle
 R_{\alpha\beta,k'}\\
& -\frac{1}{2}\sum_{\alpha k'}
\Big(
  \langle \Phi_{\alpha, k'}|1\rangle \langle 1|\Phi_{\alpha, k+k'}\rangle 
\\ &
\qquad\quad\,+\langle \Phi_{\alpha, k'-k}|1\rangle \langle 1|\Phi_{\alpha,
    k'}\rangle 
\Big) f(\epsilon_{\alpha,k'}-\mu_L) 
\bigg] .
\end{split}
\label{I_fourier}
\end{equation}
Here, we have introduced the expectation values
\begin{align}
R_{\alpha\beta}(t)&=\langle c_\alpha^\dagger(t) c_\beta(t)\rangle_t
 =R_{\beta\alpha}^*(t) \\
&=\sum_k e^{-ik\Omega t}R_{\alpha\beta,k}.
\end{align}
The Fourier decomposition in the last line is possible because all
$R_{\alpha\beta}(t)$ are expectation values of a linear, dissipative,
periodically driven system and therefore share in the long-time limit
the time-periodicity of the driving field.
In the subspace of a single electron, $R_{\alpha\beta}$ reduces to
the density matrix in the basis of the Floquet states which has
been used to describe dissipative driven quantum systems
in Refs.~\onlinecite{Blumel1991a, Dittrich1993a, Kohler1997a, Kohler1998a,
Grifoni1998a, Hanggi2000a}.

\subsection{Master equation}
The last step towards the stationary current is to find the Fourier
coefficients $R_{\alpha\beta,k}$ at asymptotic times.  To this end, we
derive an equation of motion for the reduced density operator
$\varrho_\mathrm{wire}(t) = \mathrm{Tr}_\mathrm{leads}\,
\varrho(t)$ by reinserting Eq.~\eqref{LvN,int} into the Liouville-von
Neumann equation~\eqref{LvN}.  We use that to zeroth order in the
molecule-lead coupling the interaction-picture density operator does
not change with time,
$\tilde\varrho(t-\tau,t_0)\approx\tilde\varrho(t,t_0)$.  A
transformation back to the Schr\"odinger picture results after tracing
out the leads' degrees of freedom in the master equation
\begin{widetext}
\begin{equation}
\label{mastereq}
\dot\varrho_\mathrm{wire}(t) =
 -\frac{i}{\hbar}[H_{\rm wire}(t),\varrho_\mathrm{wire}(t)] 
 -\frac{1}{\hbar^2}\int\limits_0^{\infty} \!\!d\tau
  \mathrm{Tr}_\mathrm{leads} [H_\mathrm{wire-leads}, 
  [\widetilde H_\mathrm{wire-leads}(t-\tau,t),
  \varrho_\mathrm{wire}(t)\otimes\varrho_\mathrm{leads,eq}]]  .
\end{equation}
\end{widetext}
Since we only consider asymptotic times $t_0\to-\infty$, we have set
the upper limit in the integral to infinity.  From Eq.~\eqref{mastereq}
follows directly an equation of motion for the $R_{\alpha\beta}(t)$.
Since all the coefficients of this equation, as well as its asymptotic
solution, are $\T$-periodic, we can split it into its Fourier components.
Finally, we obtain for the $R_{\alpha\beta,k}$ the inhomogeneous set of
equations
\begin{align}
\lefteqn{\frac{i}{\hbar}(\epsilon_\alpha-\epsilon_\beta+k \hbar \Omega)R_{\alpha\beta,k}}
\label{mastereq_fourier}
\\ &=&
 \frac{\Gamma_L}{2}\sum_{k'}
    \Big( &
    \sum_{\beta'k''}
                \langle\Phi_{\beta,k'+k''}|1\rangle
                \langle 1|\Phi_{\beta',k+k''}\rangle
                R_{\alpha\beta',k'}
\nonumber \\&& {}+{} &
    \sum_{\alpha'k''}
                \langle\Phi_{\alpha',k'+k''}|1\rangle
                \langle 1|\Phi_{\alpha,k+k''}\rangle
                R_{\alpha'\beta,k'}
\nonumber \\ && {}-{} &
    \langle\Phi_{\beta,k'-k}|1\rangle
    \langle 1|\Phi_{\alpha,k'}\rangle
    f(\epsilon_{\alpha,k'}-\mu_L)
\nonumber \\ && {}-{} &
    \langle\Phi_{\beta,k'}|1\rangle
    \langle 1|\Phi_{\alpha,k'+k}\rangle
    f(\epsilon_{\beta,k'}-\mu_L)
   \Big)
\nonumber \\ &&&\hspace{-10ex} {} +
  \text{same terms with the replacement}
\nonumber\\ &&&\hspace{-8ex} {}
  {\big\{\Gamma_L, \mu_L, |1\rangle\langle 1|\big\} \rightarrow
   \big\{\Gamma_R, \mu_R, |N\rangle\langle N|\big\}}.
\nonumber
\end{align}
For a consistent Floquet description, the current formula together with
the master equation must obey class invariance.
Indeed, the simultaneous transformation with (\ref{classinvariance_k})
of both the master equation (\ref{mastereq_fourier}) and the current
formula (\ref{I_fourier}) amounts to a mere shift of summation indices and,
thus, leaves the current as a physical quantity unchanged.

For the typical parameter values used below, a large number of sidebands
contributes significantly to the Fourier decomposition of the Floquet modes
$|\Phi_{\alpha}(t)\rangle$.  Numerical convergence for the solution of the
master equation (\ref{mastereq_fourier}), however, is already obtained by just
using a few sidebands for the decomposition of $R_{\alpha\beta}(t)$.  This
keeps the numerical effort relatively small and justifies \textit{a posteriori}
the use of the Floquet representation (\ref{c_n}). Yet we are able to treat
the problem beyond a rotating-wave-approximation.

\subsection{Average current}

Equation \eqref{I_L(t)} implies that the current $I_L(t)$ obeys the
time-periodicity of the driving field. Since we consider here
excitations by a laser field, the corresponding frequency lies in the
optical or infrared spectral range. In an experiment one will thus
only be able to measure the time-average of the current.  For the net
current entering through the left contact it is given by
\begin{equation}
\label{dc_current_l}
\begin{split}
\bar{I}_L =  I^0_L = 
e\Gamma_L\sum_{\alpha k}\Big[&
\sum_{\beta k'}
 \langle \Phi_{\alpha, k'+k} |1\rangle\langle 1|\Phi_{\beta, k'} \rangle
 R_{\alpha\beta,k}
\\ &
-
\langle \Phi_{\alpha, k}|1\rangle \langle 1|\Phi_{\alpha, k}\rangle
  f(\epsilon_{\alpha,k}-\mu_L)
\Big] .
\end{split}
\end{equation}
\textit{Mutatis mutandis} we obtain for the time-averaged net current
that enters through the right contact
\begin{equation}
\label{dc_current_r}
\begin{split}
\bar{I}_R =
e\Gamma_R\sum_{\alpha k}\Big[&
\sum_{\beta k'}
 \langle \Phi_{\alpha, k'+k} |N\rangle\langle N|\Phi_{\beta, k'} \rangle
 R_{\alpha\beta,k}
\\ &
-
\langle \Phi_{\alpha, k}|N\rangle \langle N|\Phi_{\alpha, k}\rangle
  f(\epsilon_{\alpha,k}-\mu_R)
\Big] .
\end{split}
\end{equation}

Total charge conservation of the original wire-lead Hamiltonian
(\ref{wire-lead-hamiltonian}) of course requires that the charge on
the wire can only change by current flow, amounting to the continuity
equation $\dot Q_{\mathrm{wire}}(t)=I_L(t)+I_R(t)$.  Since
asymptotically, the charge on the wire obeys at most the periodic
time-dependence of the driving field, the time-average of $\dot
Q_{\mathrm{wire}}(t)$ must vanish in the long-time limit.  From the
continuity equation one then finds that $\bar I_L + \bar
I_R=0$, and we can introduce the time-averaged current
\begin{equation}
\bar I = \bar I_L = -\bar I_R.
\label{barI}
\end{equation}

For consistency, the last equation must also follow from our
expressions for the average current (\ref{dc_current_l}) and
(\ref{dc_current_r}).  In fact, this can be shown by identifying $\bar
I_L + \bar I_R$ as the sum over the right-hand sides of the master
equation (\ref{mastereq_fourier}) for $\alpha=\beta$ and $k=0$,
\begin{eqnarray}
\bar I_L + \bar I_R 
&=& \sum_\alpha\left[\frac{i}{\hbar}
    (\epsilon_\alpha-\epsilon_\beta +k\hbar\Omega)R_{\alpha\beta,k}
    \right]_{\alpha=\beta,k=0} \nonumber\\
&=& 0,\label{continuity}
\end{eqnarray}
which vanishes as expected.

\subsection{Rotating-wave approximation}

Although we can now in principle compute time-dependent currents beyond a
rotating-wave approximation (RWA), it is instructive to see under what
conditions one may employ this approximation and how it follows from the
master equation (\ref{mastereq_fourier}).  We note that from a computational
viewpoint there is no need to employ a RWA since within the present approach
the numerically costly part is the computation of the Floquet states rather
than the solution of the master equation.  Nevertheless, our motivation is
that a RWA allows for an analytical solution of the master equation to lowest
order in the
lead-wire coupling $\Gamma$.  We will use this solution below to discuss the
influence of symmetries on the $\Gamma$-dependence of the average current.

The master equation (\ref{mastereq_fourier}) can be solved approximately by
assuming that the coherent oscillations of all $R_{\alpha\beta}(t)$ are much
faster than their decay.  Then it is useful to factorize $R_{\alpha\beta}(t)$
into a rapidly oscillating part that takes the coherent dynamics into account
and a slowly decaying prefactor.  For the latter, one can derive a new master
equation with oscillating coefficients.  Under the assumption that the
coherent and the dissipative time-scales are well separated, it is possible to
replace the time-dependent coefficients by their time-average.  The remaining
master equation is generally of a simpler form than the original one.  Because
we work here already with a spectral decomposition of the master equation, we
give the equivalent line of argumentation for the Fourier coefficients
$R_{\alpha\beta,k}$.

It is clear from the master equation (\ref{mastereq_fourier}) that if
\begin{equation}
\epsilon_\alpha-\epsilon_\beta+k\hbar\Omega\gg \Gamma_{L/R} ,
\label{RWA_condition}
\end{equation}
then the corresponding $R_{\alpha\beta,k}$ emerge to be small and, thus, may
be neglected.
Under the assumption that the wire-lead couplings are weak and that
the Floquet spectrum has no degeneracies, the RWA condition (\ref{RWA_condition})
is well satisfied except for
\begin{equation}
\label{RWA_condition2}
\alpha=\beta,\quad k=0,
\end{equation}
i.e.\ when the prefactor of the l.h.s.\ of Eq.~(\ref{RWA_condition}) vanishes
exactly. This motivates the ansatz
\begin{equation}
R_{\alpha\beta,k}=P_\alpha\,\delta_{\alpha,\beta}\,\delta_{k,0} ,
\label{RWA_ansatz}
\end{equation}
which means physically that the stationary state consists of an incoherent
population of the Floquet modes.  The occupation probabilities $P_\alpha$ are
found by inserting the ansatz (\ref{RWA_ansatz}) into the master
equation (\ref{mastereq_fourier}) and read
\begin{equation}
\label{RWA_solution}
P_\alpha = \frac{\sum_k\big[w_{\alpha,k}^1 f(\epsilon_{\alpha,k}-\mu_L)
                           +w_{\alpha,k}^N f(\epsilon_{\alpha,k}-\mu_R)\big]}
                {\sum_k(w_{\alpha,k}^1+w_{\alpha,k}^N)} .
\end{equation}
Thus, the populations are determined by an average over the Fermi
functions, where the weights
\begin{align}
w_{\alpha,k}^1 &= \Gamma_L|\langle 1|\Phi_{\alpha,k}\rangle|^2 , \\
w_{\alpha,k}^N &= \Gamma_R|\langle N|\Phi_{\alpha,k}\rangle|^2 ,
\end{align}
are given by the effective coupling strengths of the $k$-th Floquet sideband
$|\Phi_{\alpha,k}\rangle$ to the corresponding lead.  The
average current (\ref{barI}) is within RWA readily evaluated to read
\begin{equation}
\label{RWAcurrent}
\begin{split}
\bar I_\mathrm{RWA}
= e\sum_{\alpha,k,k'} &
    \frac{w_{\alpha,k}^1 w_{\alpha,k'}^N}
     {\sum_{k''}(w_{\alpha,k''}^1+ w_{\alpha,k''}^N)} \\
&  \times \big[ f(\epsilon_{\alpha,k'}-\mu_R)-
                f(\epsilon_{\alpha,k}-\mu_L)\big] .
\end{split}
\end{equation}

\section{Rectification of the driving-induced current}
\label{sec:mix}

In the absence of an applied voltage, i.e.\ $\mu_L=\mu_R$, the average
force on the electrons on the wire vanishes.  Nevertheless, it may
occur that the molecule rectifies the laser-induced oscillating
electron motion and consequently a non-zero dc current through the
wire is established.  In this section we investigate such ratchet
currents in molecular wires.

As a working model we consider a molecule consisting of a donor and an
acceptor site and $N-2$ sites in between (cf.\ Fig.~\ref{fig:wire}).  Each of
the $N$ sites is coupled to its nearest neighbors by a hopping matrix
elements $\Delta$.  The laser field renders each level oscillating in time with
a position dependent amplitude.  The corresponding time-dependent wire Hamiltonian reads
\begin{equation}
\begin{split}
H_{nn'}(t)= & 
-\Delta(\delta_{n,n'+1}+\delta_{n+1,n'}) \\
&+ \left[ E_n - a(t)\,x_n\right]\delta_{nn'},
\label{wirehamiltonian}
\end{split}
\end{equation}
where $x_n=(N+1-2n)/2$ is the scaled position of site $|n\rangle$, the
energy $a(t)$ equals the electron charge multiplied by the
time-dependent electrical field of the laser and the distance between
two neighboring sites. The energies of the donor and the acceptor
orbitals are assumed to be at the level of the chemical potentials of
the attached leads, $E_1=E_N=\mu_L=\mu_R$.  The bridge levels $E_n$,
$n=2,\dots,N-1$, lie $E_B$ above the chemical potential, as sketched in
Fig.~\ref{fig:wire}.  Later, we will also study the modified bridge
sketched in Fig.~\ref{fig:ratchetwire}, below.  We remark that for the
sake of simplicity, intra-atomic dipole excitations are neglected
within our model Hamiltonian.
 
In all numerical studies, we will use the hopping matrix element $\Delta$ as
the energy unit; in a realistic wire molecule, $\Delta$ is of the order
$0.1\,{\rm eV}$.  Thus, our chosen wire-lead hopping rate $\Gamma=0.1
\Delta/\hbar$ yields $e\Gamma=2.56\times10^{-5}$\,Amp\`ere and
$\Omega=3\Delta/\hbar$ corresponds to a laser frequency in the infrared.  Note
that for a typical distance of {$5$\AA} between two neighboring sites, a
driving amplitude $A=\Delta$ is equivalent to an electrical field strength of
$2\times10^6\,\mathrm{V/cm}$.

\subsection{Symmetry}
\label{subsec:symmetry}

It is known from the study of deterministically rocked periodic
potentials \cite{Flach2000a} and of overdamped classical Brownian
motion \cite{Reimann2001a} that the symmetry of the equations of
motion may rule out any non-zero average current at asymptotic times.
Thus, before starting to compute ratchet currents, let us first
analyze what kind of symmetries may prevent the sought-after effect.
Apart from the principle interest, such situations with vanishing
average current are also of computational relevance since they allow
to test numerical implementations.

The current formula (\ref{I_fourier}) and the master equation
(\ref{mastereq_fourier}) contain, besides Fermi factors, the overlap of the
Floquet states with the donor and the acceptor orbitals $|1\rangle$ and
$|N\rangle$.  Therefore, we focus on symmetries that relate these two.  If we
choose the origin of the position space at the center of the wire, it is the
parity transformation $\mathcal P:x\to -x$ that exchanges the donor with the
acceptor, $|1\rangle\leftrightarrow|N\rangle$.  Since we deal here with Floquet
states $|\Phi_{\alpha}(t)\rangle$, respectively with their Fourier
coefficients $|\Phi_{\alpha,k}\rangle$, we must take also into account the
time~$t$.
This allows for a variety of generalizations of the parity that differ by the
accompanying transformation of the time coordinate.  For a Hamiltonian of the
structure~(\ref{wirehamiltonian}), two symmetries come to mind:
$a(t)=-a(t+\pi/\Omega)$ and $a(t)=-a(-t)$.
Both are present in the case of a purely harmonic driving, i.e.\
$a(t)\propto\sin(\Omega t)$. We shall derive their consequences for
the Floquet states in the Appendix~\ref{app:symmetry} and shall only argue here why
they yield a vanishing average current within the present perturbative
approach.

\subsubsection{Generalized parity}

As a first case, we investigate a driving field that obeys
$a(t)=-a(t+\pi/\Omega)$.  Then, the wire Hamiltonian (\ref{wirehamiltonian}) is
invariant under the so-called generalized parity transformation
\begin{equation}
\label{S_GP}
\mathcal{S}_\mathrm{GP}:(x,t)\to(-x,t+\pi/\Omega).
\end{equation}
Consequently, the Floquet states are either even or odd under this
transformation, i.e.\ they fulfill the relation (\ref{app:phiGP}), which reduces
in the tight-binding limit to
\begin{equation}
\label{GPsymmetry}
\langle 1|\Phi_{\alpha,k}\rangle
= \sigma_\alpha (-1)^k\langle N|\Phi_{\alpha,k}\rangle
\end{equation}
where $\sigma_\alpha=\pm1$, according the generalized parity of the Floquet
state $|\Phi_\alpha(t)\rangle$.

The average current $\bar I$ is defined in Eq.~(\ref{barI}) by the current
formulae (\ref{dc_current_l}) and (\ref{dc_current_r}) together with the
master equation (\ref{mastereq_fourier}).  We apply now the symmetry relation
(\ref{GPsymmetry}) to them in order to interchange donor state~$|1\rangle$ and
acceptor state~$|N\rangle$.  In addition we substitute in both the master
equation and the current formulae $R_{\alpha\beta,k}$ by $\widetilde{R}_{\alpha\beta,k} =
\sigma_\alpha\sigma_\beta(-1)^k R_{\alpha\beta,k}$.  The result is
that the new expressions for the current, including the master equation, are
identical to the original ones except for the fact that $\bar I_L,\Gamma_L$
and $\bar I_R,\Gamma_R$ are now interchanged (recall that we consider the case
$\mu_L=\mu_R$).  Therefore, we can conclude that
\begin{equation}
\label{balance}
\frac{\bar I_L}{\Gamma_L}=\frac{\bar I_R}{\Gamma_R} ,
\end{equation}
which yields together with the continuity relation (\ref{continuity}) a
vanishing average current $\bar I=0$.

\subsubsection{Time-reversal parity}

A further symmetry is present if the driving is an odd function of time,
$a(t)=-a(-t)$.  Then, as detailed in the Appendix~\ref{app:symmetry}, the
Floquet eigenvalue equation (\ref{floquet_hamiltonian}) is invariant under the
time-reversal parity
\begin{equation}
\label{S_TP}
\mathcal{S}_\mathrm{TP}:(\Phi,x,t)\to(\Phi^*,-x,-t),
\end{equation}
i.e.\ the usual parity together with by time-reversal and complex conjugation
of the Floquet states $\Phi$.  The consequence for the Floquet states is the
symmetry relation (\ref{app:phiTP}) which reads for a tight-binding system
\begin{equation}
\label{TPsymmetry}
\langle 1|\Phi_{\alpha,k}\rangle
= \langle N|\Phi_{\alpha,k}\rangle^*
= \langle \Phi_{\alpha,k}|N\rangle .
\end{equation}
Inserting this into the current formulae (\ref{dc_current_l}) and
(\ref{dc_current_r}) would yield, if $R_{\alpha\beta,k}$ were real, again the
balance condition (\ref{balance}) and, thus, a vanishing average current.
However, the
$R_{\alpha\beta,k}$ are in general only real for $\Gamma_L=\Gamma_R=0$,
i.e.\ for very weak coupling such that the condition (\ref{RWA_condition}) for
the applicability of the rotating-wave approximation holds.  Then, the solution
of the master equation is dominated by the RWA
solution~(\ref{RWA_ansatz}), which is real.
In the general case, the solution of the master equation
(\ref{mastereq_fourier}) is however complex and consequently the symmetry
(\ref{TPsymmetry}) does not inhibit a ratchet effect.  Still we can conclude
from the fact that within the RWA the average current vanishes, that $\bar I$ is of the
order $\Gamma^2$ for $\Gamma\to 0$, while it is of the order $\Gamma$ for
broken time-reversal symmetry.

\subsection{Rectification from harmonic mixing}

\begin{figure}
\includegraphics[width=.95\columnwidth]{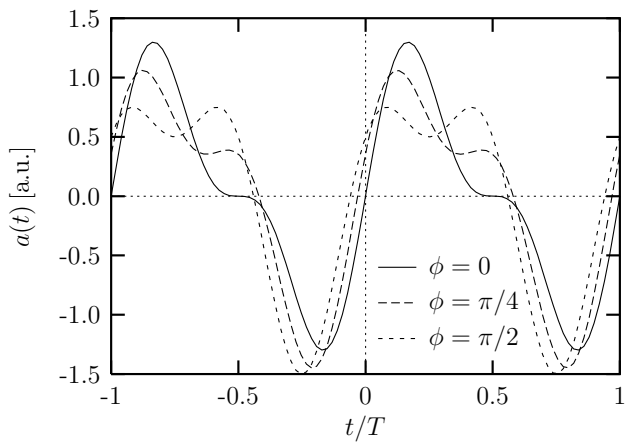}
\caption{\label{fig:mix_field}
Shape of the harmonic mixing field $a(t)$ in Eq.~(\ref{mixing}) for $A_1=2A_2$
for different phase shifts $\phi$.  For $\phi=0$, the field changes its sign
for $t\to -t$ which amounts to the time-reversal parity (\ref{S_TP}).}
\end{figure}

The symmetry analysis in Sec.~\ref{subsec:symmetry} explains that a symmetric
bridge like the one sketched in Fig.~\ref{fig:wire} will not result in a
average current if the driving is purely harmonic since a non-zero value is
forbidden by the generalized parity (\ref{S_GP}).  A simple way to break the
time-reversal part of this symmetry is to add a second harmonic to the driving
field, i.e., a contribution with twice the fundamental frequency $\Omega$,
such that it is of the form
\begin{equation}
\label{mixing}
a(t) = A_1\sin(\Omega t) + A_2\sin(2\Omega t+\phi) ,
\end{equation}
as sketched in Fig.~\ref{fig:mix_field}.  While now shifting the time $t$ by a
half period $\pi/\Omega$ changes the sign of the fundamental frequency
contribution, the second harmonic is left unchanged.  The generalized parity
is therefore broken and we find generally a non-vanishing average current.

\begin{figure}
\includegraphics[width=.95\columnwidth]{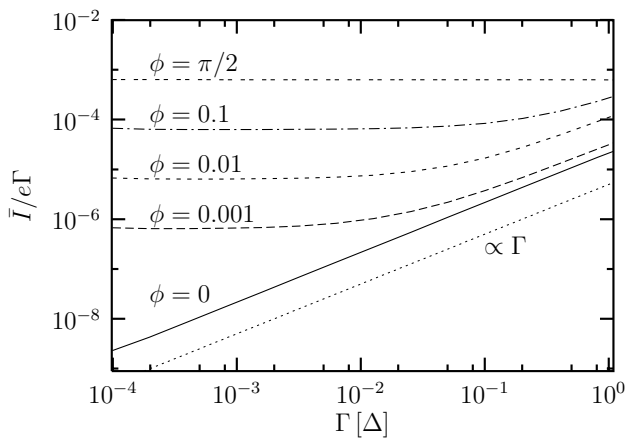}
\caption{\label{fig:mix_gamma}
Average current response to the harmonic mixing signal with amplitudes
$A_1=2A_2=\Delta$, as a function
of the coupling strength for different phase shifts $\phi$.
The remaining parameters are $\Omega=10\Delta/\hbar$, $E_\mathrm{B}=5\Delta$,
$k_BT=0.25\Delta$.
The dotted line is proportional to $\Gamma$; it represents a current which is
proportional to $\Gamma^2$.
}
\end{figure}%
The phase shift $\phi$ plays here a subtle role.  For $\phi=0$ (or equivalently
any multiple of $\pi$) the time-reversal parity is still present.  Thus,
according to the symmetry considerations above, the current vanishes within
the rotating-wave approximation.  However, as discussed above, we
expect beyond RWA for small coupling a current $\bar I\propto\Gamma^2$. 
Figure~\ref{fig:mix_gamma} confirms this prediction. Yet one
observes that already a small deviation from
$\phi=0$ is sufficient to restore the usual weak coupling behavior, namely a
current which is proportional to the coupling strength $\Gamma$.

\begin{figure}
\includegraphics[width=.95\columnwidth]{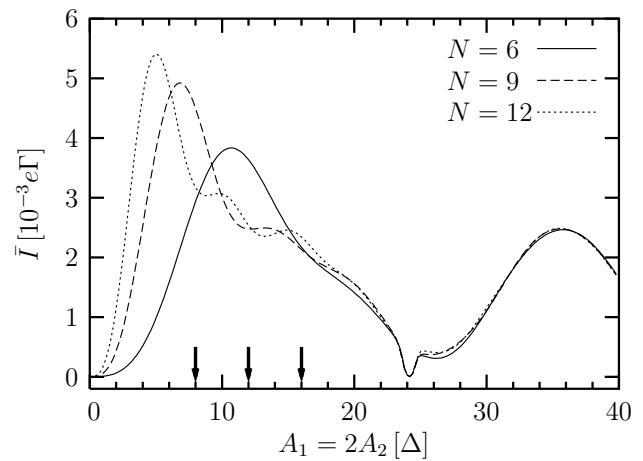}
\caption{\label{fig:mix_a}
Average current response to the harmonic mixing signal (\ref{mixing})
for $\Omega=10\,\Delta/\hbar$ and phase $\phi=\pi/2$.
The wire-lead coupling strength is $\Gamma=0.1\,\Delta$, the temperature
$k_\mathrm{B}T=0.25\Delta$, and the bridge height $E_\mathrm{B}=5\,\Delta$.
The arrows indicate the driving amplitudes used in Fig.~\ref{fig:mix_N}. 
}
\end{figure}

\begin{figure}
\includegraphics[width=.95\columnwidth]{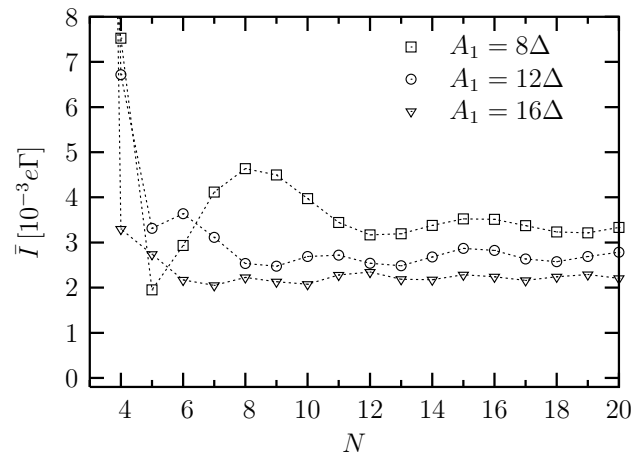}
\caption{\label{fig:mix_N}
Length dependence of the average current for harmonic mixing with phase
$\phi=\pi/2$ for different driving amplitudes; 
the ratio of the driving amplitudes is fixed by $A_1=2A_2$.
The other parameters are as in Fig.~\ref{fig:mix_a};
the dotted lines serve as a guide to the eye.
}
\end{figure}

The average current for such a harmonic mixing situation is depicted in
Fig.~\ref{fig:mix_a}.  For large driving amplitudes, it is
essentially independent of the wire length and, thus, a wire that consists
of only a few orbitals, mimics the behavior of an infinite tight-binding
system.  Figure \ref{fig:mix_N} shows the length dependence of the average
current for different driving strengths.  The current saturates as a function of
the length at a non-zero value.  The convergence depends on the driving
amplitude and is typically reached once the number of sites exceeds a value of
$N\approx 10$.  For low driving amplitudes the current response is more
sensitive to the wire length.
%
\subsection{Rectification in ratchet-like structures}

\begin{figure}
\includegraphics[width=\columnwidth]{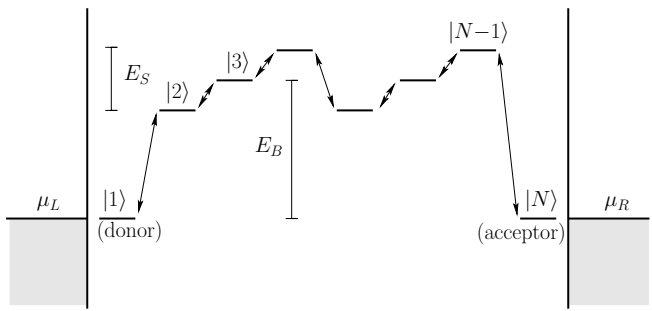}
\caption{\label{fig:ratchetwire}
Level structure of the wire ratchet with $N=8$ atomic sites, i.e.,
$N_g=2$ asymmetric molecular groups.
The bridge levels are $E_B$ above the donor and acceptor levels and are
shifted by $\pm E_S/2$.
}
\end{figure}

A second possibility to realize a finite DC current is to preserve the
symmetries of the time-dependent part of the Hamiltonian by employing a
driving field of the form
\begin{equation}
a(t)=A\sin(\Omega t),
\end{equation}
while making the level structure of the molecule asymmetric.  An
example is shown in Fig.~\ref{fig:ratchetwire}.\cite{Lehmann2002b,
on_Lehmann2002b}  In this molecular wire model, the inner wire states are
arranged in $N_g$ groups of three, i.e.\ $N-2=3N_g$. The levels in each such
group are shifted by $\pm E_S/2$, forming an asymmetric saw-tooth like
structure. 
 
\begin{figure}
\includegraphics[width=0.97\columnwidth]{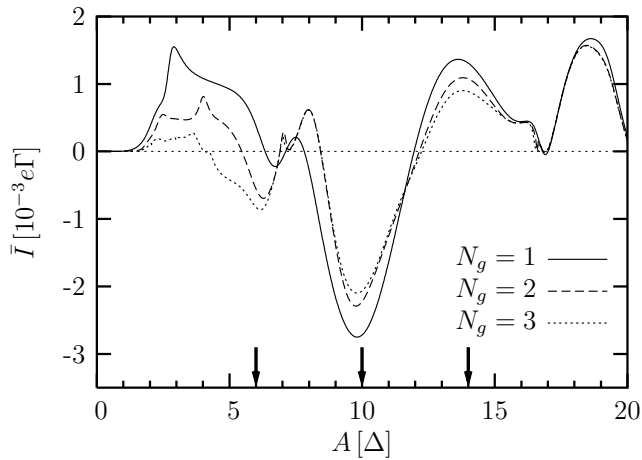}
\caption{\label{fig:I-F}
Time-averaged current through a molecular wire that consists of $N_g$
bridge units as a function of the driving strength $A$.  The bridge
parameters are $E_B=10 \Delta$, $E_S=\Delta$, the driving frequency is
$\Omega=3\Delta/\hbar$, the coupling to the leads is chosen as
$\Gamma_L=\Gamma_R=0.1 \Delta/\hbar $, and the temperature
is $k_B T=0.25 \Delta$.
The arrows indicate the driving amplitudes used in Fig.~\ref{fig:I-N}.
}
\end{figure}

\begin{figure}
\includegraphics[width=0.97\columnwidth]{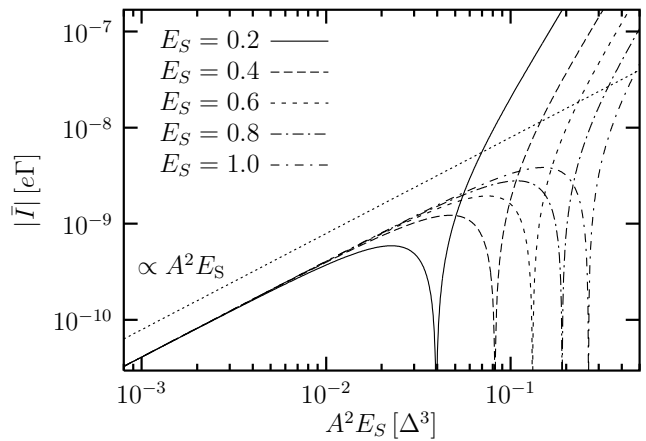}
\caption{\label{fig:I-EcA}
Absolute value of the time-averaged current in a ratchet-like structure with
$N_g=1$ as a function of $A^2 E_S$ demonstrating the
proportionality to $A^2 E_S$ for small driving amplitudes.
All other parameters are as in Fig.~\ref{fig:I-F}.
At the dips on the right-hand side, the current $\bar I$ changes its sign.
}
\end{figure}

\begin{figure}
\includegraphics[width=0.97\columnwidth]{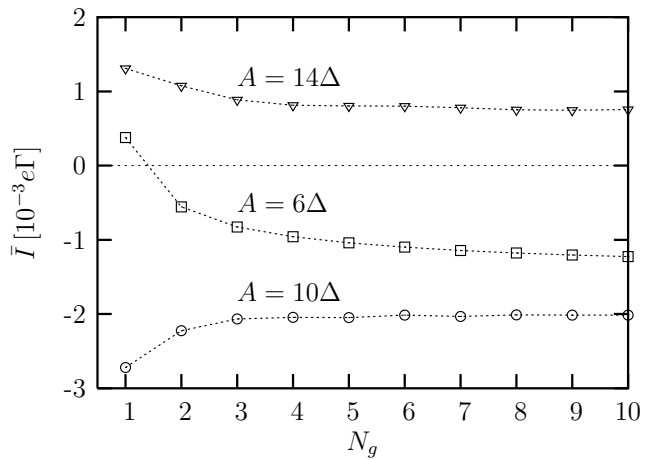}
\caption{\label{fig:I-N}
  Time-averaged current as a function of the number of bridge units $N_g$,
 $N=3N_g+2$, for the laser amplitudes indicated in Fig.~\ref{fig:I-F}.  All
 other parameters are as in Fig.~\ref{fig:I-F}.  The connecting lines serve as
 a guide to the eye.  }
\end{figure}

Figure~\ref{fig:I-F} shows for this model the stationary time-averaged
current $\bar{I}$ as a function of the driving amplitude $A$.
In the limit of a very weak laser field, we find $\bar{I}\propto A^2 E_S$,
as can be seen from Fig.~\ref{fig:I-EcA}.
This behavior is expected from symmetry
considerations: On one hand, the asymptotic current must be independent of any
initial phase of the driving field and therefore is an even function of the
field amplitude $A$.  On the other hand, $\bar I$ vanishes for zero step size
$E_S$ since then both parity symmetries are restored.  The $A^2$-dependence
indicates that the ratchet effect can only be obtained from a treatment \textit{beyond
linear response}.
For strong laser fields, we find that $\bar{I}$ is almost independent
of the wire length.  If the driving is moderately strong, $\bar{I}$
depends in a short wire sensitively on the driving amplitude $A$ and
the number of asymmetric molecular groups $N_g$; even the sign of the
current may change with $N_g$, i.e.\ we find a current reversal as a
function of the wire length.  For long wires that comprise five or
more wire units (corresponding to $17$ or more sites), the average
current becomes again length-independent, as can be observed in
Fig.~\ref{fig:I-N}.  This identifies the current reversal as
a finite size effect.  

\begin{figure}
\includegraphics[width=0.97\columnwidth]{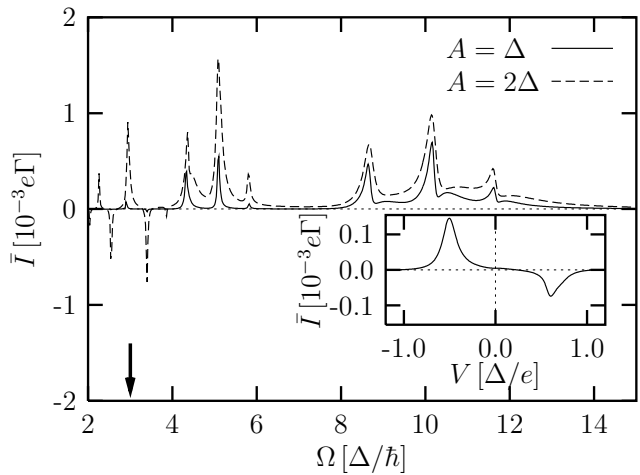}
\caption{\label{fig:I-omega}
Time-averaged current as a function of the angular
driving frequency $\Omega$ for $N_g=1$.  All other parameters
are as in Fig.~\ref{fig:I-F}. The inset displays the dependence of the average
current on an externally applied static voltage $V$, which we assume here to
drop solely along the molecule. The driving frequency and amplitude are
$\Omega=3\Delta/\hbar$ (cf. arrow in main panel) and $A=\Delta$, respectively.
}
\end{figure}

Figure~\ref{fig:I-omega} depicts the average current \textit{vs.}\ the driving
frequency~$\Omega$, exhibiting resonance peaks as a striking feature.
Comparison with the quasienergy spectrum reveals that each peak
corresponds to a non-linear resonance between the donor/acceptor
and a bridge orbital.
While the broader peaks at $\hbar\Omega\approx E_B=10 \Delta$ match the
1:1 resonance (i.e.\ the driving frequency equals the energy difference),
one can identify the sharp peaks for $\hbar\Omega\lesssim 7 \Delta$ as
multi-photon transitions.
Owing to the broken spatial symmetry of the wire, one expects an asymmetric
current-voltage characteristic. This is indeed the case as depicted
with the inset of Fig.~\ref{fig:I-omega}.

\section{Conclusions}

With this work we have detailed our recently presented approach
\cite{Lehmann2002b} for the computation of the current through a
time-dependent nanostructure.  The Floquet solutions of the isolated wire provide
a well-adapted basis set that keeps the numerical effort for the solution of
the master equation relatively low.  This allows an efficient theoretical
treatment that is feasible even for long wires in combination with strong laser
fields.

With this formalism we have investigated the possibility to rectify
with a molecular wire an oscillating external force brought about by
laser radiation, thereby inducing a non-vanishing average current without
any net bias.  A general requirement for this effect is the absence of
any reflection symmetry, even in a generalized sense.  A most
significant difference between ``true'' ratchets and molecular wires
studied here is that the latter lack the strict spatial periodicity
owing to their finite length.
However, as demonstrated above, already relatively short wires that
consist of approximately 5 to 10 units can mimic the behavior of an
infinite ratchet.  If the wire is even shorter, we find under certain
conditions a current reversal as a function of the wire length, i.e.\ 
even the sign of the current may change.  This demonstrates that the
physics of a coherent quantum ratchet is richer than the one of its units,
i.e.\ the combination of coherently coupled wire units, the driving, and
the dissipation resulting from the coupling to leads bears new intriguing effects.
A quantitative analysis of a tight-binding model has demonstrated that the
resulting currents lie in the range of $10^{-9}$\,Amp\`ere and, thus, can be
measured with today's techniques.

An alternative experimental realization of the presented results is possible
in semiconductor heterostructures, where, instead of a molecule, coherently
coupled quantum dots \cite{Blick1996a} form the central system.  A suitable
radiation source that matches the frequency scales in this case must operate
in the microwave spectral range.

\section{Acknowledgement}
We appreciate helpful discussions with S\'ebastien Camalet, Igor Goychuk,
Gert-Ludwig Ingold, and Gerhard Schmid.
This work has been supported by Sonderforschungsbereich 486 of the Deutsche
For\-schungs\-ge\-mein\-schaft and by the Volkswagen-Stiftung under grant
No.~I/77~217.

\appendix
\section{Parity of a system under driving by a dipole field}
\label{app:symmetry}

Although we describe in this work the molecule within a tight-binding
approximation, it is more convenient to study its symmetries as a function
of a continuous position and to regard the discrete sites as a special case.
Let us first consider a Hamiltonian that is an even function of $x$ and, thus,
is invariant under the parity transformation $\mathcal{P}:x\to-x$.
Then, its eigenfunctions $\varphi_\alpha$ can be divided into two classes: even and
odd ones, according to the sign in $\varphi_\alpha(x)=\pm\varphi_\alpha(-x)$.

Adding a periodically time-dependent dipole force $xa(t)$ to such a
Hamiltonian evidently breaks parity symmetry since $\mathcal{P}$ changes the
sign of the interaction with the radiation.
In a Floquet description, however, we deal with states that are functions
of both position and time---we work in the extended space $\{x,t\}$.
Instead of the stationary Schr\"odinger equation, we address the eigenvalue
problem
\begin{equation}
\label{floquet_app}
\mathcal{H}(x,t)\,\Phi(x,t) = \epsilon\,\Phi(x,t)
\end{equation}
with the so-called Floquet Hamiltonian given by
\begin{equation}
\label{H+xf}
\mathcal{H}(t)=H_0(x) + xa(t)-i\hbar\frac{\partial}{\partial t} ,
\end{equation}
where we assume a symmetric static part, $H_0(x)=H_0(-x)$.
Our aim is now to generalize the notion of parity to the extended
space $\{x,t\}$ such that the overall transformation leaves the Floquet
equation (\ref{floquet_app}) invariant.
This can be achieved if the shape of the driving $a(t)$ is such that
an additional time transformation ``repairs'' the acquired minus sign.
We consider two types of transformation:
generalized parity and time-reversal parity.
Both occur for purely harmonic driving, $a(t)=\sin(\Omega t)$.
In the following two sections we derive their consequences
for the Fourier coefficients
\begin{equation}
\label{fourier}
\Phi_k(x) = \frac{1}{T}\int_0^T \!\!\!dt\, e^{ik\Omega t}\Phi(x,t)
\end{equation}
of a Floquet states $\Phi(x,t)$.
 
\subsection{Generalized parity}
\label{app:GPsymmetry}
 
It has been noted \cite{Grossmann1991a, Grossmann1991b, Peres1991a} that a
Floquet Hamiltonian of the form (\ref{H+xf}) with $a(t)=\sin(\Omega t)$ may
possess degenerate quasienergies due to its symmetry under the so-called
generalized parity transformation
\begin{equation}
\label{app:S_GP}
\mathcal{S}_\mathrm{GP}:\quad(x,t)\to(-x,t+\pi/\Omega) ,
\end{equation}
which consists of spatial parity plus a time shift by half a driving period.
This symmetry is present in the Floquet Hamiltonian (\ref{H+xf}), if
the driving field obeys $a(t)=-a(t+\pi/\Omega)$, since
then $\mathcal{S}_\mathrm{GP}$ leaves the Floquet equation invariant.
Owing to $\mathcal S_\mathrm{GP}^2=1$, we find that
the corresponding Floquet states are either even or odd,
$\mathcal{S}_\mathrm{GP}\Phi(x,t)=\Phi(-x,t+\pi/\Omega)=\pm\Phi(x,t)$.
Consequently, the Fourier coefficients (\ref{fourier}) obey the relation
\begin{equation}
\label{app:phiGP}
\Phi_k(x)=\pm(-1)^k\Phi_k(-x) .
\end{equation}
 
\subsection{Time-inversion parity}
\label{app:TPsymmetry}

A further symmetry is found if $a$ is an odd function of time,
$a(t)=-a(-t)$.  Then, time inversion transforms the Floquet
Hamiltonian (\ref{H+xf}) into its complex conjugate so that the
corresponding symmetry is given by the anti-linear transformation
\begin{equation}
\label{app:S_TP}
\mathcal{S}_\mathrm{TP}:\quad(\Phi,x,t)\to(\Phi^*,-x,-t).
\end{equation}
This transformation represents a further generalization of the parity
$\mathcal P$; we will refer to it as \textit{time-inversion parity} since
in the literature the term generalized parity is mostly used
in the context of the transformation (\ref{app:S_GP}).

Let us now assume that that the Floquet Hamiltonian is invariant
under the transformation (\ref{app:S_TP}), $\mathcal H(x,t)=\mathcal H^*(-x,-t)$,
and that $\Phi(x,t)$ is a Floquet state, i.e., a solution of the eigenvalue
equation (\ref{floquet_app}) with quasienergy $\epsilon$. Then,
$\Phi^*(-x,-t)$ is also a Floquet state with the same quasienergy.
In the absence of any degeneracy, both Floquet states must be identical and,
thus, we find as a consequence of the time-inversion parity
$\mathcal{S}_\mathrm{TP}$ that $\Phi(x,t)=\Phi^*(-x,-t)$.
This is not nessecarily the case in the presence of degeneracies, but
then we are able to choose linear combinations of the (degenerate)
Floquet states which fulfill the same symmetry relation.
Again we are interested in the Fourier decomposition~(\ref{fourier})
and obtain
\begin{equation}
\label{app:phiTP}
\Phi_k(x)=\Phi_k^*(-x) .
\end{equation}

The time-inversion discussed here can be generalized by an additional
time-shift to read $t\to t_0-t$.  Then, we find by the same line of
argumentation that $\Phi_k(x)$ and $\Phi_k^*(-x)$ differ at most by a phase
factor.  However, for convenience one may choose already from the start the
origin of the time axis such that $t_0=0$.


\end{document}